\documentclass[sigconf, nonacm, table]{acmart}

\usepackage{multirow}
\usepackage{pifont}

\AtBeginDocument{%
  }

\setcopyright{none}
\copyrightyear{2018}
\acmYear{2018}
\acmDOI{XXXXXXX.XXXXXXX}

\acmConference[Conference acronym 'XX]{Make sure to enter the correct
  conference title from your rights confirmation email}{June 03--05,
  2018}{Woodstock, NY}

\acmISBN{978-1-4503-XXXX-X/2018/06}

\newcommand{\sysName}{\textit{RecXplore}}

\begin{document}

\title{What Matters in LLM-Based Feature Extractor for Recommender? \\ A Systematic Analysis of Prompts, Models, and Adaptation}

\author{Kainan Shi}
\affiliation{
  \institution{Xi'an Jiaotong University}
    \city{Xi'an} 
  \country{China} 
}
\email{kainanshi@stu.xjtu.edu.cn}

\author{Peilin Zhou}
\affiliation{%
  \institution{Hong Kong University of Science and Technology (Guangzhou)}
  \city{Guangzhou} 
  \country{China}
  }
\email{zhoupalin@gmail.com}

\author{Ge Wang}
\affiliation{%
  \institution{Xi'an Jiaotong University}
    \city{Xi'an} 
  \country{China} 
}
\email{gewang@xjtu.edu.cn}

\author{Han Ding}
\affiliation{
  \institution{Xi'an Jiaotong University}
    \city{Xi'an} 
  \country{China} 
}
\email{dinghan@xjtu.edu.cn}

\author{Fei Wang}
\authornote{Corresponding author.} %
\affiliation{%
  \institution{Xi'an Jiaotong University}
    \city{Xi'an} 
  \country{China} 
}
\email{feynmanw@xjtu.edu.cn}

\begin{abstract}
Using Large Language Models (LLMs) to generate semantic features has been demonstrated as a powerful paradigm for enhancing Sequential Recommender Systems (SRS).  
This paradigm typically involves three stages: converting item metadata into textual prompts, extracting semantic features using LLMs, and adapting the resulting representations for downstream recommendation models. 
However, existing approaches differ widely in prompting strategies, architectural designs, and adaptation methods, making it difficult to compare them fairly or identify the key drivers of performance.
To address this, we propose \sysName, a modular diagnostic framework that systematically disentangles the LLM-as-feature-extractor pipeline into four components: data processing, semantic feature extraction, feature adaptation, and sequential modeling.
\sysName~revisits and systematizes established methods under a unified and controlled setup, enabling fair comparisons across modules in isolation and revealing the impact of individual design choices.
Experiments on four public datasets show that assembling the best practices from each module—without exhaustive search—achieves up to 18.7\% relative improvement in NDCG@5 and 15.1\% in HR@5 over strong baselines. 
These findings highlight the value of modular analysis in identifying effective design patterns and fostering more standardized, replicable research in LLM-enhanced recommendation.
Code will be released upon acceptance.
\end{abstract}

\ccsdesc[500]{Information systems~Recommender systems}

\keywords{Recommender Systems, Large Language Models, Sequential Recommendation, Feature Extraction}

\maketitle
\section{Introduction}\label{sec:introduction_new}

\begin{figure}[t]
    \centering
    \includegraphics[width=1\linewidth]{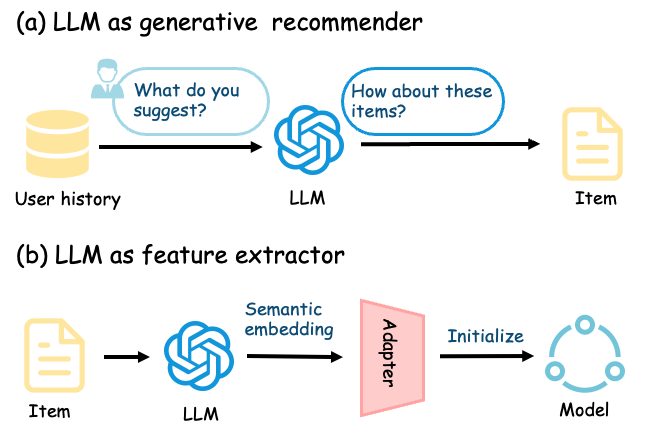}
    \caption{Two representative paradigms of applying LLMs in recommender systems.}
    \label{fig:paradigms}
\end{figure}

Recently, large language models (LLMs), known for their strong semantic understanding capabilities, have been increasingly integrated into recommender systems, demonstrating substantial potential in user modeling~\cite{user_modeling}, item representation~\cite{item_representation}, and reasoning tasks \cite{reasoning}. Among the various LLM-based approaches, two paradigms have received the most attention: generation-based recommenders (Figure~\ref{fig:paradigms} (a)) that place LLMs at the core, and representation-based methods that leverage LLMs as feature extractors to enhance traditional models (Figure~\ref{fig:paradigms} (b)). The latter offers key advantages: semantic embeddings can be precomputed offline to eliminate real-time inference latency and can be seamlessly integrated into existing architectures, making it more practical for real-world deployment \cite{wu2024survey}. Yet, how best to design and deploy this paradigm remains an open question, motivating a closer look at its internal mechanisms.

The LLM-as-feature-extractor paradigm typically follows three stages: first, transforming structured or unstructured item attributes (e.g., title, category, brand) into natural language prompts; second, feeding the prompt into an LLM to extract semantic embeddings through a designated aggregation strategy; and third, adapting the high-dimensional LLM outputs (e.g., 4096-dim vectors from LLaMA~\cite{llama}) to a compact representation compatible with downstream recommender models. Finally, the adapted representations are fed into sequential recommenders such as SASRec~\cite{kang2018self} for training. Existing studies have explored a variety of techniques for each stage. For prompt construction, \cite{hou2022towards,hou2023learning} adopt attributes flatten strategy, while others incorporate keyword extraction~\cite{lyu2023llm}, summarization~\cite{zhang2024embsum}, or knowledge-enhanced prompting~\cite{zheng2024harnessing,xu2024prompting}. In the text encoding stage, aggregation strategies range from mean pooling to last-token or special-token representations~\cite{zhang2024notellm,chen2024hllm}. For parameter tuning, earlier works~\cite{hou2022towards,zhang2024id} mostly froze the LLM, whereas recent studies~\cite{liu2025llmemb} show that lightweight fine-tuning methods, such as supervised contrastive learning, can further boost performance. Feature adaptation designs include linear projection~\cite{sheng2024language}, multilayer perceptrons (MLPs)~\cite{yuan2023go}, and mixture-of-expert (MoE)~\cite{hou2022towards} networks.

Despite these efforts, existing studies often investigate isolated components or propose single-model designs, resulting in limited comparability across methods. More importantly, multiple design decisions—such as prompt construction, representation learning, feature adaptation, and ID integration—are often tightly coupled within monolithic pipelines, making it difficult to attribute empirical gains to specific choices. 
The absence of a unified and controlled analytical perspective therefore hinders a principled understanding of what truly drives performance in the LLM-as-feature-extractor paradigm.
This raises a natural question: \textbf{How do individual design choices within the LLM-as-feature-extractor pipeline \emph{affect} recommendation performance under controlled comparisons?} 
Moreover, can stronger results be achieved by systematically identifying and combining effective practices, rather than resorting to increasingly complex architectures?

To address this question, we propose \textbf{\sysName}, a \emph{modular diagnostic framework} for systematic analysis of the LLM-as-feature-extractor pipeline. Rather than treating the pipeline as an inseparable whole, \sysName~ explicitly factorizes it into four core modules: \emph{Data Processing}, \emph{Feature Extraction}, \emph{Feature Adaptation}, and \emph{Sequential Modeling}, enabling each design dimension to be examined in isolation under consistent experimental settings.  
Through controlled experiments on four widely-used public datasets, \sysName~yields several practically actionable insights:

\textbf{First}, simple attribute flattening emerges as a robust prompting strategy, while excessive prompt engineering often introduces noise and degrades performance. 
\textbf{Second}, a two-stage LLM adaptation pipeline—continued pretraining (CPT) followed by supervised fine-tuning (SFT)—consistently produces more transferable semantic representations, with mean pooling outperforming alternative aggregation strategies. 
\textbf{Third}, effective feature adaptation depends on both dimensionality reduction and expressive alignment, where a hybrid design combining principal component analysis (PCA) and mixture-of-experts (MoE) achieves the best performance. 
\textbf{Finally}, when LLM-derived semantic embeddings are sufficiently expressive, traditional ID embeddings provide marginal benefit, making direct replacement the most effective integration strategy.

Building on these insights, we instantiate the identified design principles into the LLM-as-feature-extractor pipeline, which consistently outperforms strong baselines across all datasets and evaluation metrics, achieving up to a \textbf{15.1\%} and \textbf{18.7\%} relative improvement in HR@5 and NDCG@5, respectively. Notably, these gains are achieved without introducing additional architectural complexity, highlighting the practical value of systematic design-space disentanglement and modular diagnosis.
Our main contributions are summarized as follows:

\begin{itemize}
\item We introduce \sysName, a controlled modular diagnostic framework that factorizes the LLM-as-feature-extractor pipeline and enables isolated evaluation of key design choices under a unified experimental protocol.
\item We systematically study major design dimensions, including prompting, LLM adaptation, feature compression, and ID integration, across multiple public datasets, yielding reproducible evidence and actionable guidelines for building LLM-enhanced sequential recommenders.
\item We show that the best-performing configuration distilled from our analysis—without architectural over-engineering or exhaustive search—consistently outperforms strong LLM-based baselines across datasets and evaluation metrics.
\end{itemize}

\section{Related Work} \label{sec:related-work}

\begin{figure*}[!t]
    \centering
    \includegraphics[width=1\linewidth]{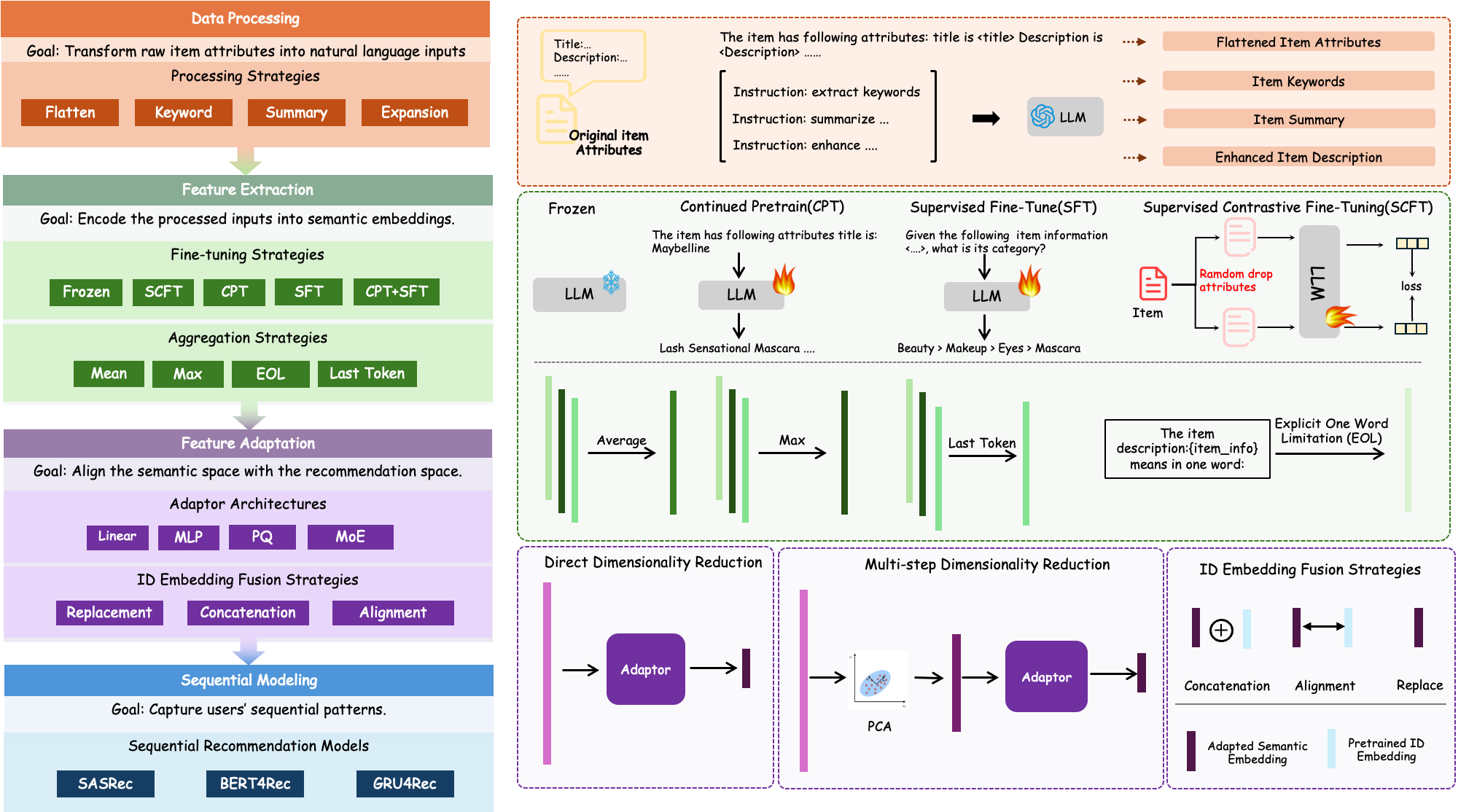}
    \caption{Overview of the proposed \textbf{\sysName} framework, which comprises four modules: \emph{Data Processing}, \emph{Feature Extraction}, \emph{Feature Adaptation}, and \emph{Sequential Modeling}. The figure illustrates the data flow from item attributes to final recommendations. For each module, we systematically examine representative design choices and their impact on recommendation performance.}
    \label{fig:framework}
\end{figure*}
\subsection{Sequential Recommendation}

Sequential recommendation aims to predict the next item a user will interact with based on their behavioral history. Early methods used Markov chains~\cite{he2016fusing} to model short-term dependencies. With deep learning, RNN-based models such as GRU4Rec~\cite{hidasi2015session} and CNN-based models like NextItNet~\cite{yuan2019simple} and Caser~\cite{tang2018personalized} were proposed to better capture sequential patterns. Transformer-based models, including SASRec~\cite{kang2018self} and BERT4Rec~\cite{sun2019bert4rec}, further improved performance by modeling long-range dependencies. Recent advances introduce Mamba-based architectures~\cite{liu2024mamba4rec, su2024mlsa4rec, wang2024echomamba4rec} that offer linear inference complexity for long sequences, and diffusion-based models~\cite{li2023diffurec, liu2023diffusion, ma2024plug} that enhance recommendation by modeling uncertainty through generative denoising processes.
However, despite their success, these models typically ignore the rich semantic information contained in item attributes such as titles, descriptions, and categories, which can provide valuable complementary signals beyond interaction IDs.

\subsection{LLM for Recommendation}


Recent research has explored two primary paradigms for integrating LLMs into recommendation pipelines. The first paradigm is LLM-centric, in which user behaviors and item attributes are converted into natural language descriptions and processed by a pre-trained LLM to directly generate recommendation outputs~\cite{geng2022recommendation, bao2023tallrec, liu2023chatgpt}. This approach treats recommendation as a language modeling task and leverages the generative strengths of LLMs. The second paradigm explores the use of LLMs as feature extractors to enhance recommender systems, focusing on different components such as prompt design~\cite{hou2022towards,zhang2024embsum,xu2024prompting}, aggregation strategies~\cite{zhang2024notellm,chen2024hllm}, or adaptation modules~\cite{yuan2023go,hou2023learning,liu2025llmemb}. While these studies demonstrate the potential of LLM-enhanced representations, they typically investigate specific techniques in isolation or under inconsistent setups, making it difficult to draw general conclusions about what design choices matter and why. This fragmentation has been noted in recent surveys, which underscore the need for standardized evaluation~\cite{wu2024survey, xi2024towards}.

In contrast, our work introduces \sysName, a unified and modular framework that systematically decomposes the LLM-as-feature-extractor pipeline into four core modules. This design enables controlled comparisons across design dimensions and datasets, allowing us to derive actionable principles for building effective LLM-enhanced recommender systems.


\section{Problem Formulation}\label{sec:problem}
We study the problem of enhancing sequential recommendation by leveraging large language models (LLMs) as feature extractors. Formally, given a user's interaction sequence $\mathcal{S}_u = [v_1, v_2, ..., v_{t-1}]$, the task is to predict the next item $v_t$ that the user is most likely to interact with. Each item $v_i$ is associated with structured or unstructured metadata $\mathcal{A}_{v_i}$ (e.g., \texttt{title}, \texttt{brand}, \texttt{category}), which we transform into a textual prompt $\mathcal{P}_{v_i}$. This prompt is fed into a pretrained LLM to generate a high-dimensional semantic embedding $\mathbf{e}_{v_i}^{\text{LLM}}$, which is then adapted into a compact representation $\mathbf{z}_{v_i}$ for use in downstream sequential models (e.g., SASRec). The key challenge lies in designing and coordinating the components of this pipeline—prompt construction, LLM aggregation, feature adaptation, and ID fusion—so that the resulting representations are both semantically rich and computationally efficient. This motivates a modular formulation, where each stage can be independently studied and optimized, as instantiated in our \sysName{} framework.

\section{RecXplore Framework} \label{sec:framework}

In this section, we present \sysName, a modular diagnostic framework for systematic analysis of the LLM-as-feature-extractor paradigm in sequential recommendation. 
As shown in Figure~\ref{fig:framework}, \sysName~ factorizes the pipeline into four modules: \textbf{Data Processing} (Sec.~\ref{data_processing}), \textbf{Feature Extraction} (Sec.~\ref{feature_extraction}), \textbf{Feature Adaptation} (Sec.~\ref{feature_adaptation}), and \textbf{Sequential Modeling} (Sec.~\ref{sequential_model}). 
Under a controlled and reproducible setup, \sysName~ isolates and evaluates representative design choices within each module, enabling fine-grained comparisons and actionable insights into what matters for LLM-based recommendation.

\subsection{Data Processing Module}
\label{data_processing}


This module transforms raw item attributes (e.g., title, brand, category, price) into natural language prompts for LLM encoding. To analyze how input formulation affects downstream performance, we evaluate four representative strategies: \textit{Attributes Flatten}, \textit{Keyword Extraction}, \textit{Summarization}, and \textit{Knowledge Expansion}. 

\begin{figure}[t]
    \centering
    \includegraphics[width=1\linewidth]{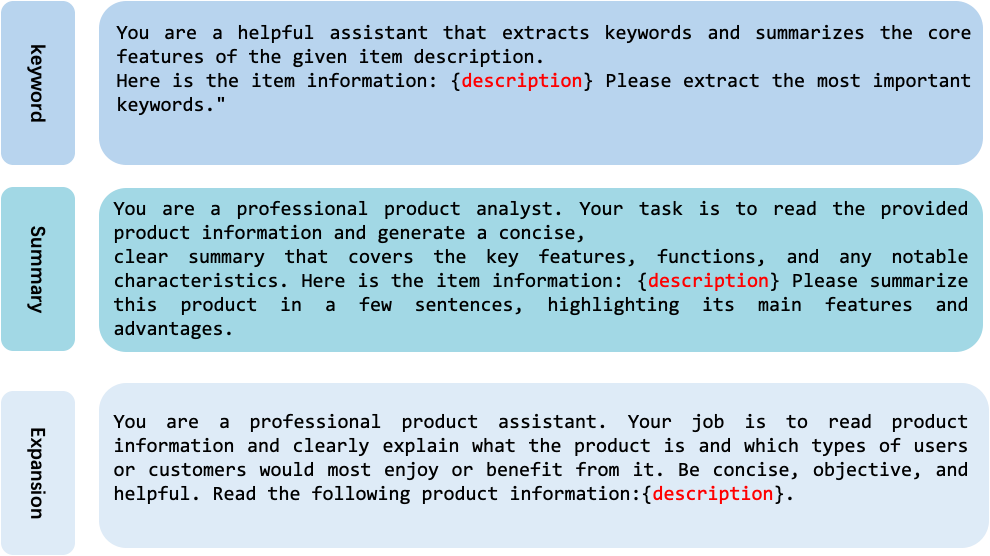}
    \caption{Instruction templates for Keyword Extraction, Summarization, and Expansion. The placeholder \texttt{{description}} denotes the flattened item attributes.}
    \label{fig:prompt_example}
\end{figure}
The first, \textit{Attributes Flatten}, is a rule-based baseline that concatenates structured fields using fixed templates (e.g., ``Brand: Nike; Category: Running Shoes; Description: Lightweight and breathable design''), offering a simple yet effective prompt format without extra computation.  
The remaining three are LLM-augmented variants that leverage an auxiliary language model (e.g., GPT-4o~\cite{achiam2023gpt}) for enriching the prompt: \textit{Keyword Extraction} retains only the most salient tokens, \textit{Summarization} compresses verbose descriptions into concise summaries~\cite{zhang2024embsum}, and \textit{Knowledge Expansion} injects external commonsense or factual knowledge~\cite{xu2024prompting,zheng2024harnessing}.  
Figure~\ref{fig:prompt_example} presents the specific instruction templates for these strategies.

\subsection{Feature Extraction Module}
\label{feature_extraction}

This module encodes processed prompts into high-dimensional semantic embeddings using a large language model (LLM).
In practice, the choice of LLM backbone (e.g., encoder-only vs.\ decoder-only architectures and model scale) can affect both embedding quality and deployment cost.
In this work, we treat the LLM backbone as a controlled factor and fix it to a representative open-source model, \textsc{LLaMA2-7B}, for reproducibility; a comparative study of alternative backbones is provided in Appendix~\ref{sec:appendix_featureExtractors}.
Under this controlled setting, we focus on two key design dimensions that most directly govern the expressiveness and transferability of item embeddings:
(1) how the LLM is fine-tuned, and (2) how token-level outputs are aggregated into item-level embeddings.


\subsubsection{LLM Fine-tuning Strategy.}
We investigate four representative fine-tuning strategies for adapting the LLM to the recommendation task: Continued Pre-training (CPT)~\cite{yildiz2024investigating}, Supervised Fine-tuning (SFT)~\cite{harada2025massive}, Supervised Contrastive Fine-tuning (SCFT)~\cite{liu2025llmemb}, and the combined CPT+SFT pipeline.  
CPT performs continued pretraining on unlabeled item texts generated via the flattening of structured attributes (e.g., brand, title, category), helping the LLM better align with domain-specific distributions.  
SFT leverages QA-style supervision, where the model is prompted with partial item descriptions (e.g., title and brand) and asked to predict missing attributes such as the category, thus injecting task-relevant semantic cues.  
CPT+SFT integrates the above two forms of supervision sequentially to enhance both general domain alignment and task-specific adaptation. 
SCFT applies random field dropout to an item's textual attributes to create two augmented views of the same item, which are treated as positives in a contrastive objective.

\subsubsection{Feature Aggregation Strategy.}
Feature aggregation determines how to convert the LLM’s output—a sequence of token embeddings—into a single vector representing the item. We evaluate four representative methods: \textit{Mean Pooling}, which averages the token hidden states; \textit{Max Pooling}, which takes the element-wise maximum; \textit{Last Token}, which uses the hidden state of the final token; and \textit{Explicit One-word Limitation (EOL)}~\cite{zhang2024simple}, which prompts the LLM to summarize the input into a single word, using that word's embedding as the final representation. 

\subsection{Feature Adaptation Module}
\label{feature_adaptation}
This module bridges the gap between high-dimensional LLM embeddings and the low-dimensional input space required by sequential recommenders. We explore two key design aspects: (1) adaptation architectures that reduce dimensionality, either directly or via multi-step schemes (e.g., PCA followed by learnable adapters); and (2) fusion strategies that integrate semantic and ID-based embeddings, allowing the model to leverage both rich item semantics and collaborative signals.

\subsubsection{Adaptation Architectures.}
\label{sec:adaptation_arch}

To reduce the high dimensionality of LLM embeddings and align them with downstream recommendation models, we evaluate two adaptation paradigms: Direct Dimensionality Reduction (DDR) and Multi-step Dimensionality Reduction (MDR) (see Figure~\ref{fig:framework}).
DDR applies a learnable adaptor to compress embeddings in a single step. We consider four representative adaptor architectures:  
(1) Linear Projection, a simple fully connected layer that performs a linear transformation;  
(2) Multilayer Perceptron (MLP), a stack of non-linear layers that increases expressiveness;  
(3) Product Quantization (PQ)~\cite{jegou2010product}, which discretizes the embedding space into multiple subspaces for compact encoding; and  
(4) Mixture-of-Experts (MoE)~\cite{shazeer2017outrageously}, where a gating network dynamically activates a subset of expert MLPs whose outputs are aggregated.
MDR first uses Principal Component Analysis (PCA) as an unsupervised preprocessor to reduce embedding dimensionality, followed by a learnable adaptor (e.g., MLP or MoE) for task-specific alignment. This design improves parameter efficiency while retaining semantic fidelity.

\subsubsection{Fusion with ID Embeddings.}
\label{fusion}
To examine whether semantic representations can benefit from collaborative signals encoded in traditional ID embeddings, we evaluate three integration strategies: (1) Replacement, where semantic embeddings substitute original ID embeddings; (2) Concatenation, which merges the two into a unified vector; and (3) Alignment, which introduces an auxiliary contrastive loss to align adapted semantic embeddings with pretrained ID embeddings from a recommendation model.

\subsection{Sequential Modeling Module.}
\label{sequential_model}
This module serves as the recommendation backbone for modeling user behavior sequences. To evaluate the generalizability of our design choices, we adopt three representative sequential recommendation models: GRU4Rec~\cite{hidasi2015session}, BERT4Rec~\cite{sun2019bert4rec}, and SASRec~\cite{kang2018self}. Each model consumes the adapted item embeddings as input to learn user preferences and generate top-$K$ recommendations.

\subsection{Training and Inference}
\subsubsection{Training.}

We investigate two training modes, distinguished by whether the LLM parameters are updated. 
In \emph{Frozen LLM} (single-stage training), the LLM is kept fixed and we optimize only the Adaptation Module and the downstream SASRec using the recommendation loss (e.g., cross-entropy). 
In \emph{Fine-tuned LLM} (two-stage training), we first adapt the LLM on recommendation data via parameter-efficient fine-tuning (PEFT; e.g., LoRA) to inject domain knowledge, and then freeze the adapted LLM to generate item embeddings for training the Adaptation Module and SASRec.
\subsubsection{Inference.}
\label{inference}
To ensure the low-latency responses required for industrial applications, all item semantic embeddings are pre-computed and cached offline. During online inference, the system only needs to perform a forward pass through the lightweight Adaptation Module and the Sequential Module. This completely avoids real-time calls to the LLM, thus guaranteeing efficient recommendation services.
A detailed analysis of the computational overhead and latency trade-offs is provided in Sec.~\ref{sec:efficiency}.

\section{Experiments} \label{sec:experiments_new}

\begin{table*}[!t]
\small
\setlength{\tabcolsep}{2pt}
\centering
\caption{Performance comparison across different data processing (RQ1), feature aggregation (RQ2), and LLM fine-tuning strategies (RQ3). \textbf{Bold} and \underline{underline} denote the best and second-best results, respectively; ${}^*$ indicates statistical significance ($p<0.05$) compared to the second-best result.}
\resizebox{\textwidth}{!}{
\begin{tabular}{lc|cc|cc|cc|cc}
\toprule
\multirow{2}{*}{\textbf{Strategy}} & & \multicolumn{2}{c|}{\textbf{Beauty}} & \multicolumn{2}{c|}{\textbf{Games}} & \multicolumn{2}{c|}{\textbf{Fashion}} & \multicolumn{2}{c}{\textbf{Steam}} \\
\cmidrule(lr){3-4} \cmidrule(lr){5-6} \cmidrule(lr){7-8} \cmidrule(lr){9-10}
 & & \textbf{H@5/H@10} & \textbf{N@5/N@10} & \textbf{H@5/H@10} & \textbf{N@5/N@10} & \textbf{H@5/H@10} & \textbf{N@5/N@10} & \textbf{H@5/H@10} & \textbf{N@5/N@10} \\
\midrule
\multicolumn{9}{l}{\textit{Impact of data processing strategies (RQ1).}} \\
\midrule
Flatten & & \textbf{0.4282*/0.5525*} & \textbf{0.3110*/0.3511*} & \underline{0.5442/0.7019} & \underline{0.3850/0.4362} & \underline{0.5121/0.5832} & \underline{0.4574/0.4803} & \textbf{0.5592*/0.7106*} & \textbf{0.4043*}/\underline{0.4469} \\

Keyword & & 0.4210/0.5443 & 0.3052/0.3450 & \textbf{0.5508*/0.7038*} & \textbf{0.3928*/0.4425*} & 0.5017/0.5659 & 0.4443/0.4650 & 0.5372/0.6989 & 0.3968/0.4391 \\

Summary & & 0.4179/0.5418 & 0.3017/0.3418 & 0.5275/0.6827 & 0.3741/0.4244 & \textbf{0.5184*/0.5860*} & \textbf{0.4599*/0.4816*} & 0.5495/0.7041 & 0.3993/0.4412 \\

Expansion & & \underline{0.4269/0.5515} & \underline{0.3093/0.3495} & 0.5430/0.6990 & 0.3846/0.4352 & 0.4971/0.5628 & 0.4394/0.4606 & \underline{0.5568/0.7086} & \underline{0.4029}/\textbf{0.4479*} \\
\midrule
\multicolumn{9}{l}{\textit{Impact of feature aggregation strategies (RQ2).}} \\
\midrule
EOL & & \underline{0.4035/0.5224} & \underline{0.2926/0.3310} & \underline{0.5339}/\underline{0.6914} & 0.3745/0.4256 & 0.4977/0.5692 & \underline{0.4327/0.4557} & \underline{0.5502/0.7028} & \underline{0.3949/0.4444} \\

Last Token & & 0.3920/0.5166 & 0.2806/0.3209 & \underline{0.5339}/0.6907 & \underline{0.3752/0.4260} & \underline{0.5031/0.5746} & 0.4361/0.4591 & 0.5393/0.6950 & 0.3860/0.4365 \\

Max Pooling & & 0.1801/0.2695 & 0.1264/0.1550 & 0.4296/0.5962 & 0.2936/0.3474 & 0.4085/0.4709 & 0.3362/0.3564 & 0.4511/0.6153 & 0.3020/0.3552 \\

Mean Pooling & & \textbf{0.4282*/0.5525*} & \textbf{0.3110*/0.3511*} & \textbf{0.5442*/0.7019*} & \textbf{0.3850*/0.4362*} & \textbf{0.5121*/0.5832*} & \textbf{0.4574*/0.4803*} & \textbf{0.5592*/0.7106*} & \textbf{0.4043*/0.4469*} \\
\midrule
\multicolumn{9}{l}{\textit{Impact of LLM fine-tuning strategies (RQ3).}} \\
\midrule
Frozen & & 0.4282/0.5525 & 0.3110/0.3511 & 0.5442/0.7019 & 0.3850/0.4362 & 0.5121/\underline{0.5832} & 0.4574/0.4803 & 0.5592/0.7106 & 0.4043/0.4469 \\

SCFT & & 0.4641/0.5705 & 0.3558/0.3901 & 0.6093/0.7443 & 0.4475/0.4905 & \underline{0.5241}/0.5740 & \textbf{0.4819*/0.4980*} & \underline{0.5964/0.7432} & \underline{0.4383/0.4859} \\

CPT & & \underline{0.4803/0.5903} & \textbf{0.3679*}/\underline{0.4034} & 0.5800/0.7261 & 0.4180/0.4654 & 0.5197/0.5823 & 0.4628/0.4830 & 0.5800/0.7380 & 0.4173/0.4686 \\

SFT & & 0.4769/0.5886 & \underline{0.3623}/0.3984 & \underline{0.6108/0.7477} & \underline{0.4518/0.4967} & 0.5212/0.5837 & 0.4664/0.4865 & 0.5929/0.7416 & 0.4328/0.4811 \\

CPT+SFT & & \textbf{0.4812*/0.5952*} & 0.3633/\textbf{0.4057*} & \textbf{0.6140*/0.7511*} & \textbf{0.4569*/0.4988*} & \textbf{0.5270*/0.5915*} & \underline{0.4680/0.4887} & \textbf{0.6056*/0.7496*} & \textbf{0.4427*/0.4895*} \\
\bottomrule
\end{tabular}
}
\label{tab:rq123}
\end{table*}

We conduct systematic experiments based on the proposed \sysName~ framework to analyze how design choices across different modules affect recommendation performance. Our study is guided by the following research questions:

\begin{itemize}
    \item\textbf{RQ1}  What are the specific impacts of different data processing strategies on overall performance?
    \item\textbf{RQ2}  {What is the impact of different feature aggregation strategies on the final recommendation performance?}
    \item\textbf{RQ3} {What is the impact of different LLM fine-tuning strategies on recommendation quality?}
    \item\textbf{RQ4}  What is the impact of different feature adaptation strategies on the final recommendation performance?
    \item\textbf{RQ5}  Can pre-trained item ID embeddings bring significant additional performance gains on top of the semantic embeddings?
    \item\textbf{RQ6}  Can the optimal combination of components, distilled from our systematic decoupled analysis, outperform the state-of-the-art methods, and by what margin?
\end{itemize}


\subsection{Experimental Settings}

\subsubsection{Dataset.}
{We conduct experiments on four  real-world datasets, i.e., the Steam dataset~\cite{martin_bustos_roman_2022} and three categories from Amazon product reviews~\cite{he2016ups}: Beauty, Fashion, and Games. These datasets span different product domains and user behavior patterns, serving as common benchmarks for evaluating sequential recommendation models. To ensure consistency, we adopt the data pre-processing method proposed in LLMEmb~\cite{liu2025llmemb} for all datasets.}

\subsubsection{Evaluation Metrics.}

 Following previous work~\cite{kang2018self}, we use Hit Rate (HR@K) and Normalized Discounted Cumulative Gain (NDCG@K) as our evaluation metrics, with K set to {5, 10}. In the evaluation, each expected recommended item in the test set is paired with 100 randomly sampled uninteracted items to calculate the metrics. To reduce the randomness of the results, all experiments are conducted three times with different random seeds (42, 43, 44), and we report the average metrics. As the standard deviation across all runs was consistently $\leq$ 0.002, we omit it from the tables for brevity.
 Additionally, to validate the performance improvements, we perform paired t-tests between the best and second-best results in our experimental tables.

\begin{table*}[!t]
\small
\setlength{\tabcolsep}{2pt}
\centering
\caption{Performance comparison across different adapter architectures, with and without PCA pre-processing (RQ4). \textbf{Bold} and \underline{underline} denote the best and second-best results, respectively; ${}^*$ indicates statistical significance ($p<0.05$) compared to the second-best result.}
\resizebox{\textwidth}{!}{
\begin{tabular}{lc|cc|cc|cc|cc}
\toprule
& & \multicolumn{2}{c|}{\textbf{Beauty}} & \multicolumn{2}{c|}{\textbf{Games}} & \multicolumn{2}{c|}{\textbf{Fashion}} & \multicolumn{2}{c}{\textbf{Steam}} \\
\cmidrule(lr){3-4} \cmidrule(lr){5-6} \cmidrule(lr){7-8} \cmidrule(lr){9-10}
\textbf{Adapter} & \textbf{PCA} & \textbf{H@5/H@10} & \textbf{N@5/N@10} & \textbf{H@5/H@10} & \textbf{N@5/N@10} & \textbf{H@5/H@10} & \textbf{N@5/N@10} & \textbf{H@5/H@10} & \textbf{N@5/N@10} \\
\midrule
\multirow{2}{*}{\textbf{Linear}} 
& \ding{55} & 0.4812/0.5952 & 0.3633/0.4057 & 0.6140/0.7511 & 0.4569/0.4988 & 0.5270/0.5915 & 0.4680/0.4887 & \underline{0.6056}/0.7496 & 0.4427/\underline{0.4895} \\
& \ding{51} & 0.4750/0.5831 & 0.3681/0.4029 & 0.6038/0.7369 & 0.4459/0.4891 & \underline{0.5489/0.6088} & \textbf{0.5105*}/\underline{0.5255} & 0.5907/0.7338 & 0.4310/0.4775 \\

\midrule
\multirow{2}{*}{\textbf{MLP}}
& \ding{55} & 0.4641/0.5814 & 0.3479/0.3859 & 0.6174/0.7563 & 0.4558/0.5009 & 0.5285/0.5993 & 0.4678/0.4906 & 0.5942/0.7466 & 0.4307/0.4802 \\
& \ding{51} & \underline{0.5053/0.6135} & 0.3908/\underline{0.4288} & \underline{0.6423/0.7641} & \underline{0.4836/0.5268} & 0.5423/0.5959 & 0.4965/0.5138 & 0.5991/0.7495 & \underline{0.4371}/0.4860 \\

\midrule
\multirow{2}{*}{\textbf{PQ}}
& \ding{55} & 0.3905/0.5008 & 0.2869/0.3225 & 0.5282/0.6738 & 0.3825/0.4297 & 0.4745/0.5309 & 0.4108/0.4322 & 0.5185/0.6748 & 0.3695/0.4201 \\
& \ding{51} & 0.3753/0.4734 & 0.2868/0.3185 & 0.4878/0.6219 & 0.3524/0.3958 & 0.5054/0.5547 & 0.4656/0.4815 & 0.4870/0.6262 & 0.3520/0.3971 \\

\midrule
\multirow{2}{*}{\textbf{MoE}}
& \ding{55} & \textbf{0.5104*/0.6155*} & \textbf{0.3978*/0.4308*} & 0.6355/0.7624 & 0.4803/0.5216 & 0.5302/0.5953 & 0.4675/0.4886 & 0.6027/\underline{0.7511} & 0.4318/0.4813 \\
& \ding{51} & 0.5066/0.6053 & \underline{0.3957}/0.4268 & \textbf{0.6464*/0.7675*} & \textbf{0.4896*/0.5289*} & \textbf{0.5544*/0.6112*} & \underline{0.5088}/\textbf{0.5282*} & \textbf{0.6227*/0.7683*} & \textbf{0.4612*/0.5062*} \\
\bottomrule
\end{tabular}
}
\label{tab:ablation_adaptation}
\end{table*}

\subsubsection{Base Configuration.}
The base configuration integrates the most straightforward and representative design choices from each module and serves as the reference setting for all subsequent decoupled analyses.
Specifically, we adopt Attribute Flatten for data processing, use a frozen LLaMA2-7B with mean pooling for feature extraction, apply a single-layer linear projection (without PCA) for feature adaptation, and employ SASRec as the downstream backbone for sequential modeling.

\subsubsection{Implementation Details.}

All experiments were conducted on 8 NVIDIA L20 GPUs. 
We selected LLaMA2-7B~\cite{llama} as the LLM backbone, considering it a more suitable choice than state-of-the-art alternatives like OpenAI Text Embedding 3~\cite{achiam2023gpt} and Qwen Embedding 8B~\cite{yang2025qwen3}, owing to its moderate parameter size and balanced performance capabilities (see Appendix~\ref{sec:appendix_featureExtractors} for a detailed comparison).
The LLM generates 4096-dimensional embeddings, which are optionally reduced to 1536 dimensions using PCA before being projected to the final 128-dimensional space required by the SASRec backbone. For the MoE adapter, we utilize 8 linear experts controlled by a noisy softmax gating network. We apply Parameter-Efficient Fine-Tuning (PEFT) using LoRA~\cite{hu2022lora} with a rank of $r=8$ and scaling factor $\alpha=32$ across all fine-tuning strategies. In terms of data processing, we apply LLM-based data augmentation using GPT-4o~\cite{achiam2023gpt}. 
For LLM fine-tuning, we observe that full data is crucial for CPT to cover long-tail items effectively. While using only 40\% of the data for SFT yields over 95\% of the performance, we report results using the full dataset to establish the rigorous upper bound. Additionally, in Sec.~\ref{sec:main_results}, we compare our approach with several LLM-enhanced recommenders, including SAID~\cite{hu2024enhancing}, LLMESR~\cite{liu2024llm}, and LLMEmb~\cite{liu2025llmemb}, using the authors' open-source code. We perform grid search over the key hyperparameters within the ranges specified in their original papers to ensure fairness in our evaluations. Our model is optimized using AdamW with a batch size of 1024, an initial learning rate of 0.001, and a maximum of 200 epochs, utilizing an early stopping strategy with a patience of 20 epochs to prevent overfitting.


\begin{table*}[!t]
\small
\setlength{\tabcolsep}{1pt}
\centering
\caption{Performance comparison of different ID feature fusion strategies. \textbf{Bold} denotes the best result, \underline{underline} indicates the second-best, and ${}^*$ indicates statistical significance ($p<0.05$) compared to the second-best result. (RQ5)}
\resizebox{\textwidth}{!}{
\begin{tabular}{lll|cc|cc|cc|cc}
\toprule
& & & \multicolumn{2}{c|}{\textbf{Beauty}} & \multicolumn{2}{c|}{\textbf{Games}} & \multicolumn{2}{c|}{\textbf{Fashion}} & \multicolumn{2}{c}{\textbf{Steam}} \\
\cmidrule(lr){4-5} \cmidrule(lr){6-7} \cmidrule(lr){8-9} \cmidrule(lr){10-11}
\multicolumn{2}{l}{\textbf{Adapter}} & \textbf{Strategy} & \textbf{H@5/H@10} & \textbf{N@5/N@10} & \textbf{H@5/H@10} & \textbf{N@5/N@10} & \textbf{H@5/H@10} & \textbf{N@5/N@10} & \textbf{H@5/H@10} & \textbf{N@5/N@10} \\
\midrule
\multirow{3}{*}{\textbf{Linear}} 
& & - Replace & \textbf{0.4750*/0.5831*} & \textbf{0.3681*/0.4029*} & \underline{0.6038/0.7369} & \underline{0.4459/0.4891} & \underline{0.5489/0.6088} & \textbf{0.5105*/0.5255*} & 0.5907/\underline{0.7338} & 0.4310/0.4775 \\

& & - Concat & 0.4649/0.5721 & 0.3567/0.3913 & \textbf{0.6390*/0.7555*} & \textbf{0.4940*/0.5318*} & 0.5297/0.5822 & 0.4819/0.4989 & \textbf{0.5975*/0.7423*} & \textbf{0.4404*/0.4873*} \\
& & - Align & \underline{0.4741/0.5817} & \underline{0.3644/0.3992} & 0.5995/0.7367 & 0.4390/0.4835 & \textbf{0.5554*/0.6095*} & \underline{0.5075/0.5249} & \underline{0.5914}/0.7317 & \underline{0.4348/0.4804} \\
\midrule
\multirow{3}{*}{\textbf{MoE}} 
& & - Replace & \textbf{0.5066*/0.6053*} & \textbf{0.3957*/0.4268*} & \textbf{0.6464*/0.7675*} & \textbf{0.4896*/0.5289*} & \textbf{0.5544*/0.6112*} & \textbf{0.5088*/0.5282*} & \textbf{0.6227*/0.7683*} & \textbf{0.4612*/0.5062*} \\
& & - Concat & 0.4485/0.5744 & 0.3401/0.3753 & 0.6338/0.7496 & \underline{0.4893}/ \underline{0.5269} & 0.5388/0.5895 & 0.4968/0.5131 & 0.5975/0.7453 & 0.4401/0.4880 \\
& & - Align & \underline{0.5001/0.6035} & \underline{0.3885/0.4220} & \underline{0.6346/0.7564} & 0.4792/0.5187 & \underline{0.5477/0.6028} & \underline{0.4985/0.5163} & \underline{0.6219/0.7630} & \underline{0.4579/0.5038} \\
\bottomrule
\end{tabular}
}
\label{tab:ablation_fusion}
\end{table*}



\subsection{Analysis of Data Processing Strategies (RQ1)}
To assess how different data processing strategies affect recommendation performance, we fix all downstream modules—including feature extraction, adaptation, and sequential modeling—and vary only the prompt construction method. Results are reported in Table~\ref{tab:rq123}, from which we draw the following observations.

\textbf{Attribute Flatten is the most robust strategy.} It achieves the best or second-best results across datasets and metrics (e.g., 0.5592/0.7106 HR@5/HR@10 on \textit{Steam}), and yields statistically significant gains over several rewriting variants. 
This suggests that preserving the original structure and fine-grained attribute semantics is critical for item-centric feature extraction, as it maintains stable and discriminative cues that better align with user interaction patterns in sequential recommendation.

\textbf{LLM-based data augmentation is not consistently helpful.} Both compression-based rewriting (keyword extraction / summarization) and knowledge expansion often fail to improve over Flatten across datasets. 
A likely reason is that rewriting introduces an additional semantic transformation step: compression may discard discriminative attributes, while expansion may inject weakly aligned information, leading to semantic drift and less stable item representations.

\subsection{Analysis of Feature Aggregation (RQ2)}


Building on the optimal data processing strategy identified in RQ1, we evaluate different feature aggregation methods by comparing \textit{Mean Pooling} with \textit{Max Pooling}, \textit{Last Token}, and \textit{Explicit One-word Limitation (EOL)}. 
All susequent modules are fixed to base configurations to isolate the effect of aggregation. The results are reported in Table~\ref{tab:rq123} and we derive the following findings:

\textbf{Mean Pooling is the most effective aggregation strategy.}
As shown in Table~\ref{tab:rq123}, Mean Pooling achieves the best performance across all datasets and metrics.
For example, on the \textit{Games} dataset, it attains 0.7019 in HR@10 and 0.4362 in NDCG@10, significantly outperforming all alternative aggregation methods.
Similar trends are observed on \textit{Beauty}, \textit{Fashion}, and \textit{Steam}, indicating that averaging token embeddings yields stable and expressive item representations by preserving distributed semantic evidence across attributes.

\textbf{Aggregation methods that collapse token-level evidence lead to inferior performance.}
Single-token strategies such as EOL and Last Token consistently underperform Mean Pooling (e.g., 0.5502 vs.\ 0.5592 HR@5 on \textit{Steam}), suggesting an inherent information bottleneck when compressing rich item semantics into a single representation.
Max Pooling performs the worst across all datasets, with its performance on \textit{Beauty} dropping to 0.1801 HR@5 and 0.1264 NDCG@5, less than half of Mean Pooling.
This indicates that overemphasizing extreme token activations amplifies noise and rare signals that are weakly aligned with user behavior, resulting in unstable item embeddings.




\subsection{Analysis of LLM Fine-tuning (RQ3)}
Having fixed the optimal data processing and aggregation strategies identified in RQ1 and RQ2, we examine how different LLM fine-tuning strategies affect recommendation performance. 
The results are summarized in Table~\ref{tab:rq123}. We have the following observations:

\textbf{LLM adaptation consistently improves recommendation performance over the frozen baseline.}
Across all datasets, unsupervised continued pre-training (CPT), supervised fine-tuning (SFT), and supervised contrastive fine-tuning (SCFT) all yield substantial gains compared to the frozen model.
For example, on the \textit{Beauty} dataset, HR@5 improves from 0.4282 (Frozen) to 0.4803 with CPT, 0.4769 with SFT, and 0.4641 with SCFT, with similar trends observed on other datasets.
These results indicate that, despite their strong general-purpose knowledge, LLMs require domain-specific adaptation to effectively capture recommendation-oriented semantics.

\textbf{Different fine-tuning strategies exhibit complementary strengths, with CPT+SFT achieving the most consistent performance.}
While SCFT and SFT directly optimize task-level objectives and outperform the frozen baseline, they do not consistently surpass CPT, suggesting limited domain coverage when supervision is applied alone.
In contrast, the two-stage CPT+SFT pipeline consistently achieves the best results across all benchmarks (e.g., 0.4812 HR@5 on \textit{Beauty} and 0.6140 HR@5 on \textit{Games}), often with statistically significant margins.
This pattern suggests a complementary effect: CPT reshapes the representation space to better reflect domain-specific distributions, while subsequent SFT refines these representations toward task-relevant signals, resulting in more robust item embeddings than single-stage adaptation.


\subsection{Analysis of Feature Adaptation (RQ4)}




Building on the best upstream settings (i.e., \textit{Attributes Flatten}, \textit{Mean Pooling}, and \textit{CPT+SFT}), we analyze the feature adaptation layer that aligns high-dimensional LLM embeddings with the recommender latent space.
Following Section~\ref{sec:adaptation_arch}, we compare DDR (a single-step learnable adaptor) and MDR (PCA followed by a learnable adaptor) using four architectures—\textit{Linear}, \textit{MLP}, \textit{PQ}, and \textit{MoE}. Results are reported in Table~\ref{tab:ablation_adaptation}, and we make the following observations:

\textbf{Under the DDR paradigm, MoE is the most effective adaptor architecture across datasets.}
MoE achieves the best or near-best performance among all DDR variants, consistently outperforming Linear, MLP, and PQ adaptors.
In particular, MoE without PCA attains the highest performance on the \textit{Beauty} dataset (H@5 = 0.5104), while remaining competitive across all other datasets.
These results suggest that the dynamic, input-dependent routing mechanism of MoE provides a more expressive and flexible mapping from the LLM embedding space to the recommendation space than static projection-based adaptors.

\textbf{Under the MDR paradigm, PCA further enhances adaptation performance, with the strongest gains observed when combined with MoE.}
Although PCA does not universally improve all adaptors or datasets, it consistently boosts MoE on three out of four datasets, leading to the overall best-performing configurations.
For example, PCA improves MoE on \textit{Games} from 0.6355 to 0.6464 in H@5 and on \textit{Steam} from 0.6027 to 0.6227.
We also observe that PCA benefits other nonlinear adaptors such as MLP, indicating that dimensionality reduction helps remove redundancy and noise in high-dimensional LLM embeddings.

\begin{table*}[!t]
\small
\setlength{\tabcolsep}{1pt}
\centering
\caption{Performance comparison between our proposed RecXplore and baselines. \textbf{Bold} denotes the best result, \underline{underline} indicates the second-best, and ${}^*$ indicates statistical significance ($p<0.05$) compared to the strongest baseline.(RQ6)} 
\resizebox{\textwidth}{!}{
\begin{tabular}{cl|cc|cc|cc|cc}
\toprule
& & \multicolumn{2}{c|}{\textbf{Beauty}} & \multicolumn{2}{c|}{\textbf{Games}} & \multicolumn{2}{c|}{\textbf{Fashion}} & \multicolumn{2}{c}{\textbf{Steam}} \\
\cmidrule(lr){3-4} \cmidrule(lr){5-6} \cmidrule(lr){7-8} \cmidrule(lr){9-10}
\textbf{Backbone} & \textbf{Model} & \textbf{H@5/H@10} & \textbf{N@5/N@10} & \textbf{H@5/H@10} & \textbf{N@5/N@10} & \textbf{H@5/H@10} & \textbf{N@5/N@10} & \textbf{H@5/H@10} & \textbf{N@5/N@10} \\
\midrule

& - None & 0.2636/0.3639 & 0.1836/0.2160 & 0.3604/0.4906 & 0.2829/0.3149 & 0.4136/0.4779 & 0.3571/0.3773 & 0.5216/0.6861 & 0.3673/0.4207 \\
& - SAID & 0.3285/0.4201 & 0.2386/0.2653 & 0.4412/0.5325 & 0.3315/0.3785 & 0.4412/0.4974 & 0.3892/0.4193 & 0.5486/0.6925 & 0.3912/0.4386 \\
\textbf{GRU4Rec} & - LLMESR & \underline{0.4075/0.5089} & \underline{0.3045/0.3225} & \underline{0.5025/0.6137} & \underline{0.3725/0.4186} & \underline{0.4825/0.5437} & \underline{0.4286/0.4593} & \underline{0.5689/0.7098} & \underline{0.4098/0.4521} \\
& - LLMEmb & 0.3871/0.4572 & 0.2593/0.2871 & 0.4819/0.5923 & 0.3619/0.3919 & 0.4519/0.5123 & 0.4190/0.4419 & 0.5512/0.6935 & 0.3986/0.4398 \\
& - RecXlpore & \textbf{0.4403*/0.5531*} & \textbf{0.3307*/0.3670*} & \textbf{0.5786*/0.7135*} & \textbf{0.4271*/0.4709*} & \textbf{0.5139*/0.5804*} & \textbf{0.4625*/0.4839*} & \textbf{0.5973*/0.7432*} & \textbf{0.4358*/0.4832*} \\
\rowcolor{gray!20} & \textit{Improv.} & \textit{+8.0\%/+8.7\%} & \textit{+8.6\%/+13.8\%} & \textit{+15.1\%/+16.3\%} & \textit{+14.7\%/+12.5\%} & \textit{+6.5\%/+6.8\%} & \textit{+7.9\%/+5.4\%} & \textit{+5.0\%/+4.7\%} & \textit{+6.3\%/+6.9\%} \\
\midrule
& - None & 0.2867/0.3991 & 0.1998/0.2361 & 0.4165/0.5529 & 0.2955/0.3396 & 0.4102/0.4679 & 0.3683/0.3868 & 0.5165/0.6779 & 0.3674/0.4196 \\
& - SAID & 0.4125/0.5155 & 0.2986/0.3391 & 0.4925/0.5985 & 0.3625/0.4225 & 0.4486/0.5172 & 0.3925/0.4185 & 0.5425/0.6845 & 0.3886/0.4289 \\
\textbf{Bert4Rec} & - LLMESR & \underline{0.4386/0.5431} & \underline{0.3385/0.3679} & \underline{0.5386/0.6527} & \underline{0.3986/0.4546} & \underline{0.4783/0.5481} & \underline{0.4186/0.4529} & \underline{0.5625/0.7086} & \underline{0.4025/0.4412} \\
& - LLMEmb & 0.4327/0.5351 & 0.3324/0.3621 & 0.5225/0.6198 & 0.3912/0.4486 & 0.4736/0.5275 & 0.4105/0.4329 & 0.5535/0.6925 & 0.3915/0.4325 \\
& - RecXlpore & \textbf{0.4815*/0.5918*} & \textbf{0.3770*/0.4127*} & \textbf{0.6113*/0.7434*} & \textbf{0.4647*/0.5075*} & \textbf{0.5208*/0.5927*} & \textbf{0.4813*/0.4917*} & \textbf{0.6159*/0.7598*} & \textbf{0.4506*/0.4974*} \\
\rowcolor{gray!20} & \textit{Improv.} & \textit{+9.8\%/+9.0\%} & \textit{+11.4\%/+12.2\%} & \textit{+13.5\%/+13.9\%} & \textit{+16.6\%/+11.6\%} & \textit{+8.9\%/+8.1\%} & \textit{+15.0\%/+8.6\%} & \textit{+9.5\%/+7.2\%} & \textit{+12.0\%/+12.7\%} \\
\midrule

& - None & 0.3290/0.4193 & 0.2521/0.2812 & 0.5489/0.6813 & 0.3989/0.4682 & 0.4683/0.5030 & 0.4364/0.4619 & 0.5541/0.6954 & 0.4016/0.4461 \\
& - SAID & 0.4218/0.4998 & 0.3067/0.3244 & 0.5603/0.6984 & 0.4096/0.4727 & 0.4982/0.5327 & 0.4389/0.4679 & 0.5607/0.7028 & 0.4089/0.4573 \\
\textbf{SASRec} & - LLMESR & \underline{0.4511/0.5692} & \underline{0.3459/0.3741} & \underline{0.5734/0.7080} & 0.4074/0.4791 & \underline{0.5167/0.5678} & \underline{0.4635/0.4866} & \underline{0.5811/0.7353} & \underline{0.4311/0.4807} \\
& - LLMEmb & 0.4362/0.5410 & 0.3340/0.3678 & 0.5654/0.7069 & \underline{0.4126/0.4892} & 0.5096/0.5554 & 0.4710/0.4856 & 0.5739/0.7327 & 0.4302/0.4789 \\
& - RecXlpore & \textbf{0.5066*/0.6053*} & \textbf{0.3957*/0.4268*} & \textbf{0.6464*/0.7675*} & \textbf{0.4896*/0.5289*} & \textbf{0.5544*/0.6112*} & \textbf{0.5088*/0.5282*} & \textbf{0.6227*/0.7683*} & \textbf{0.4612*/0.5062*} \\
\rowcolor{gray!20} & \textit{Improv.} & \textit{+12.3\%/+6.3\%} & \textit{+14.4\%/+14.1\%} & \textit{+12.7\%/+8.4\%} & \textit{+18.7\%/+8.1\%} & \textit{+7.3\%/+7.6\%} & \textit{+8.0\%/+8.5\%} & \textit{+7.2\%/+4.5\%} & \textit{+7.0\%/+5.3\%} \\
\bottomrule
\end{tabular}
}
\label{tab:sota_comparison}
\end{table*}

\subsection{Analysis of Item ID Embeddings Fusion (RQ5)}




Having established the best practices for semantic adaptation, we examine the role of traditional ID embeddings in the \emph{LLM-as-feature-extractor} paradigm.
We consider two representative adapter regimes—Linear (lightweight) and MoE (high-capacity)—and evaluate three ID fusion strategies: Replacement, Concatenation, and Alignment.
Table~\ref{tab:ablation_fusion} shows that the effectiveness of ID fusion is strongly conditioned on adapter capacity:

\textbf{For the MoE adapter, direct replacement is consistently optimal.}
Across all datasets, replacing ID embeddings with semantic representations yields the best performance, whereas concatenation or alignment does not lead to further improvements.
This suggests that the semantic representations produced by MoE are sufficiently informative for item discrimination, making explicit ID signals largely redundant in this setting.
From an optimization perspective, semantic embeddings provide dense, attribute-level supervision: many dimensions contribute gradients for learning a task-aligned mapping, and the expressive MoE adapter can exploit this richness via conditional expert specialization.
In contrast, ID embeddings are sparse and instance-specific; injecting them alongside strong semantic features can encourage identity memorization with limited transferability and may interfere with preserving semantic structure, leading to negligible marginal gains.

\textbf{For the Linear adapter, incorporating ID embeddings remains beneficial.}
In this case, fusing ID information via concatenation or alignment consistently improves over direct replacement.
For example, on \textit{Games}, concatenation increases HR@5 from 0.6038 to 0.6390.
This indicates that a linear adapter may underfit the semantic-to-recommendation alignment; ID embeddings then provide a complementary, fast-to-learn identity signal via direct lookup, improving separability among semantically similar items.

\subsection{Putting It All Together: Comparison with Baselines (RQ6)}
\label{sec:main_results}






Based on the systematic investigations facilitated by the \sysName{} framework, we have identified the optimal design choices for each module of the \textit{LLM-as-feature-extractor} paradigm. 
By synthesizing these best practices, we instantiate an \textbf{optimized configuration} that integrates: (1) a flattened attribute concatenation strategy for item description; (2) mean pooling for prompt-level embedding aggregation; (3) a two-stage CPT+SFT scheme for LLM adaptation; (4) a PCA-enhanced MoE adapter for feature compression and transformation; and (5) a direct replacement of ID embeddings with semantic vectors for downstream sequential models.

We integrate this optimized configuration into three widely adopted sequential recommenders: GRU4Rec~\cite{hidasi2015session}, Bert4Rec~\cite{sun2019bert4rec}, and SASRec~\cite{kang2018self}. We compare its performance against three representative LLM-enhanced baselines\footnote{We focus on the feature extractor paradigm compatible with efficient retrieval. Generative models (e.g., TALLRec~\cite{bao2023tallrec}) are excluded due to their high inference latency, which prevents fair comparison in this context.}: SAID~\cite{hu2024enhancing}, LLMESR~\cite{liu2024llm}, and LLMEMB~\cite{liu2025llmemb}. Additionally, to verify generalizability across LLM architectures, we extend our evaluation by replacing the Llama-2-7B backbone with the lightweight Qwen2.5-3B.\footnote{Due to space limitations, we only report results on the Beauty dataset; trends on other datasets are consistent with these findings.} The key findings are summarized below:

\textbf{The optimized configuration consistently achieves SOTA performance.} 
As shown in Table~\ref{tab:sota_comparison}, our proposed approach outperforms all baselines across datasets and evaluation metrics. For instance, under the SASRec backbone, it achieves a substantial \textbf{18.7\%} improvement in NDCG@5 on Games and a \textbf{14.4\%} improvement in HR@5 on Beauty compared to the best-performing LLM-based baseline. This result validates that the specific combination of components identified by \sysName{} is superior to existing monolithic architectures.

\textbf{Improvements are robust across downstream architectures.} 
Crucially, the performance gains are not limited to a specific model type. We observe consistent improvements across all evaluated backbones, from the RNN-based GRU4Rec to Transformer-based models like SASRec and Bert4Rec. This demonstrates that the derived pipeline provides a model-agnostic solution for injecting semantic knowledge into sequential recommendation.

\noindent\textbf{The configuration generalizes well to smaller LLM backbones.} 
As detailed in Table~\ref{tab:cross_model_generalization}, we observe a moderate performance decrease compared to Llama-2-7B (0.4811 vs. 0.5066). This phenomenon is expected and corroborates existing findings that larger parameter scales generally yield richer semantic representations~\cite{li2025exploring}. 
However, even under this parameter constraint, our method maintains a dominant lead over competitive baselines, notably surpassing the strongest competitor (LLMESR) by \textbf{12.5\%}. 
This provides compelling evidence that the effectiveness of our approach stems primarily from the principled modular design rather than relying solely on the sheer scale of the foundation model.

\begin{table}[t] 
  \centering
  \caption{Performance comparison of different LLM backbones on the Beauty dataset, utilizing SASRec as the downstream recommender.}
  \label{tab:cross_model_generalization}
  \resizebox{0.48\textwidth}{!}{
  \begin{tabular}{c|l|cc|cc}
    \toprule
    LLM & Method & HR@5 & HR@10 & NDCG@5 & NDCG@10 \\
    \midrule
    \multirow{3}{*}{\shortstack{Llama-2-7B}} 
      & LLMESR & 0.4511 & 0.5692 & 0.3459 & 0.3741 \\
      & LLMEmb & 0.4362 & 0.5410 & 0.3340 & 0.3678 \\
      & \textbf{RecXplore} & \textbf{0.5066} & \textbf{0.6053} & \textbf{0.3957} & \textbf{0.4268} \\
    \midrule
    \multirow{3}{*}{\shortstack{Qwen2.5-3B}} 
      & LLMESR & 0.4275 & 0.5442 & 0.3237 & 0.3683 \\
      & LLMEmb & 0.4106 & 0.5184 & 0.3061 & 0.3470 \\
      & \textbf{RecXplore} & \textbf{0.4811} & \textbf{0.5835} & \textbf{0.3719} & \textbf{0.4051} \\
    \bottomrule
  \end{tabular}
  }
\end{table}


\subsection{Computational Efficiency}
\label{sec:efficiency}

To validate the practical feasibility of RecXplore, we explicitly analyze the computational trade-offs, distinguishing between online inference latency and offline training overhead.

\textbf{Online Latency.} 
A critical advantage of our framework is that it incurs \textbf{zero additional online latency} compared to standard ID-based methods. As detailed in Sec.~\ref{inference}, since all semantic embeddings are pre-computed and cached, the online item encoding step serves as a standard $O(1)$ table lookup. This design effectively decouples the adapter's architectural complexity from the inference stage. Consequently, deploying a complex MoE adapter introduces no latency penalty compared to a simple Linear adapter during real-time serving.

\textbf{Offline Overhead.} 
We evaluate the training costs on the Beauty dataset ($\approx$ 57k items) using LLaMA-7B on 8 NVIDIA L20 GPUs. 
Regarding training time, the optimal \textit{CPT+SFT} strategy requires an additional $\approx$ 45 minutes compared to the 1.5-hour baseline of the Frozen setting. This one-time offline investment is well-justified by the substantial 18.7\% performance gain reported in Sec.~\ref{sec:main_results}.
 Regarding feature generation, the MoE adapter adds negligible marginal overhead ($\approx$ 0.04s total increase for the entire dataset) relative to a Linear adapter. 
In summary, RecXplore achieves significant performance improvements with manageable offline costs and strictly no compromise on online efficiency.

\section{Conclusion and Future Work} \label{sec:conclusion}


In this work, we present \textbf{RecXplore}, a modular framework for optimizing the \textit{LLM-as-feature-extractor} paradigm in sequential recommendation. By revisiting and decoupling the pipeline into four core components (data processing, feature extraction, feature adaptation, and sequential modeling), RecXplore enables a fine-grained diagnostic analysis of each design dimension. While prior studies often focus on isolated techniques or propose monolithic solutions, they tend to overlook the broader interplay among components, limiting the understanding of what truly contributes to performance. Our framework addresses this gap by systematically comparing representative choices within each module and assembling them into an optimized configuration. Experiments show that simple combinations of best practices—such as attribute flattening, mean pooling, and CPT+SFT tuning—consistently outperform recent LLM-based baselines, with up to 18.7\% improvement in NDCG@5 and 15.1\% in HR@5. These findings highlight the practical value of revisiting and rethinking existing designs, and suggest that structured analysis and principled integration can be more effective than pursuing complex architectures alone.

In the future, we plan to explore joint optimization across modules, expand the framework to incorporate multi-modal attributes, and investigate its applicability to broader recommendation settings beyond sequential modeling.

\appendix
\begin{figure}[h!]
    \centering
    \includegraphics[width=1\linewidth]{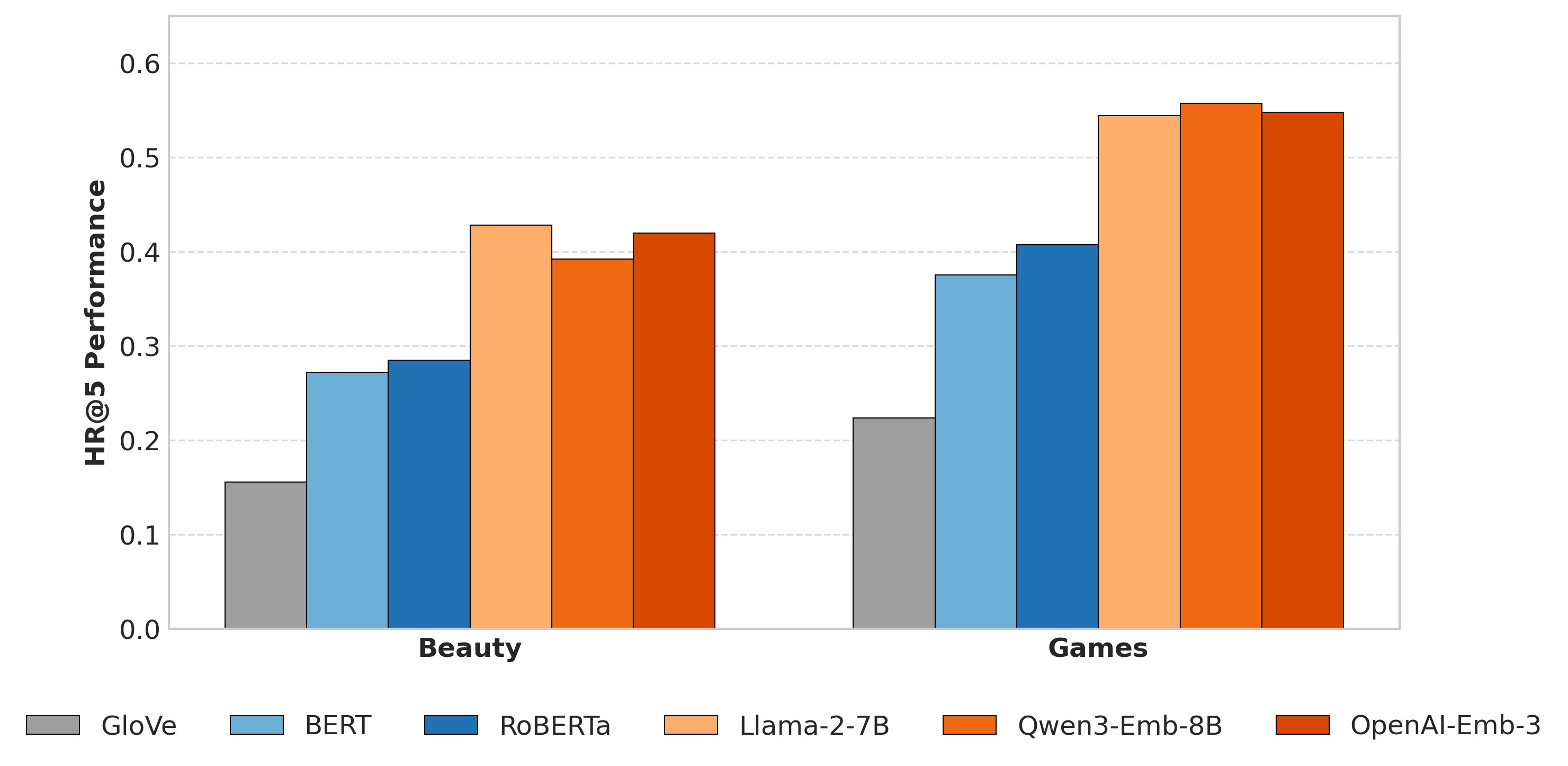}
    \caption{Performance comparison of different feature extractors on Beauty and Games datasets.}
    \label{fig:model_comparison}
\end{figure}

\begin{table}[h!]
\centering
\caption{Performance comparison of BERT and RoBERTa feature extractors on the Beauty dataset. (Best results within each model group are bolded, second-best are underlined.)}
\label{tab:bert_roberta_beauty}
\resizebox{\columnwidth}{!}{%
\begin{tabular}{@{}llcccc@{}}
\toprule
& & \multicolumn{4}{c}{\textbf{Beauty}} \\
\cmidrule(l){3-6}
\textbf{Model} & \textbf{Pooling} & \textbf{H@5} & \textbf{H@10} & \textbf{N@5} & \textbf{N@10} \\
\midrule
BERT & [CLS] & 0.1887 & 0.2758 & 0.1339 & 0.1619 \\
& First-Last Avg & \underline{0.2680} & \textbf{0.3885} & \underline{0.1795} & \underline{0.2183} \\
& Max & 0.2342 & 0.3461 & 0.1566 & 0.1926 \\
& Mean & \textbf{0.2721} & \underline{0.3881} & \textbf{0.1850} & \textbf{0.2223} \\
\midrule
RoBERTa & [CLS] & 0.1800 & 0.2628 & 0.1282 & 0.1548 \\
& First-Last Avg & \underline{0.2672} & \textbf{0.3837} & \underline{0.1808} & \underline{0.2183} \\
& Max & 0.2253 & 0.3377 & 0.1498 & 0.1860 \\
& Mean & \textbf{0.2751} & \underline{0.3812} & \textbf{0.1890} & \textbf{0.2242} \\
\bottomrule
\end{tabular}%
}
\end{table}

\section{Appendix} \label{sec:appendix}
\subsection{Additional Results on Feature Extractor Backbones}
\label{sec:appendix_featureExtractors}

\paragraph{Backbone Comparison.}
We first compare representative feature extractors from three model categories: static embeddings (GloVe), small encoder-only models (BERT and RoBERTa), and large language models (LLaMA2-7B, Qwen3-Emb-8B, and OpenAI Emb-3). 
As shown in Figure~\ref{fig:model_comparison}, LLM-based feature extractors consistently outperform both static embeddings and small encoder-only models on the \textit{Beauty} and \textit{Games} datasets. 
This result indicates that larger-capacity language models provide substantially stronger semantic representations for sequential recommendation in our evaluation setting.
Among the LLM backbones, Qwen3-Emb-8B achieves slightly better performance on \textit{Games}, while OpenAI Emb-3 remains competitive overall. 
We adopt \textbf{LLaMA2-7B} as the default backbone in our main experiments due to its strong and stable performance across domains, together with the advantage of open-source reproducibility.


\paragraph{Aggregation Strategies for Encoder-only Models.}
To further examine the generality of our findings on feature aggregation, we conduct an additional study using encoder-only models as feature extractors.
Table~\ref{tab:bert_roberta_beauty} reports the performance of BERT and RoBERTa under different aggregation strategies on the \textit{Beauty} dataset.
Consistent with our main results, mean pooling achieves the best or second-best performance across most evaluation metrics for both models.
For instance, it boosts BERT's H@5 from 0.1887 (using \texttt{[CLS]}) to 0.2721, proving that distributed token representations capture richer semantics than single-token bottlenecks.
These results suggest that the advantage of mean pooling is not limited to LLM-based feature extractors, but also generalizes to smaller encoder-only models, reinforcing the robustness of our conclusions, although their absolute performance still lags behind the larger generative models shown in Figure ~\ref{fig:model_comparison}.

\newpage


\bibliographystyle{ACM-Reference-Format}
\bibliography{sample-base}


\end{document}